\documentclass[traditabstract]{aa}

\usepackage{graphicx}
\usepackage{natbib}

\usepackage{piero-macro-private}

\newcommand{\ML}{(\texttt{ML})}
\newcommand{\W}{(\texttt{W})}

\begin{document}
\title{The XMM-Newton survey in the H-ATLAS field
\thanks{Based on observations obtained with XMM-Newton, an ESA science mission with instruments and contributions directly funded by ESA Member States and NASA.}
}

   \author{P.~Ranalli
          \inst{1,2}
          \and
          I.~Georgantopoulos\inst{1}
          \and
          A.~Corral\inst{1}
          \and
          L.~Koutoulidis\inst{1}
          \and
          M.~Rovilos\inst{1}
          \and
          F.~J.~Carrera\inst{3}
          \and
          A.~Akylas\inst{1}
          \and
          A.~Del Moro\inst{4}
          \and
          A.~Georgakakis\inst{5}
          \and
          R.~Gilli\inst{2}
          \and
          C.~Vignali\inst{6}
          }

   \institute{
     Institute for Astronomy, Astrophysics, Space Applications and
     Remote Sensing (IAASARS), National Observatory of
     Athens,  15236 Penteli, Greece; 
     \email{piero.ranalli@noa.gr} 
    \and
     INAF -- Osservatorio Astronomico di Bologna,
     via Ranzani 1, 40127 Bologna, Italy
     \and
     Instituto de F\'{\i}sica de Cantabria (CSIC-UC), 39005 Santander, Spain
    \and
    Durham University, Department of Physics, South Road, Durham DH1
    3LE, UK
    \and
     Max-Planck-Institut f\"ur extraterrestrische Physick, 85478
     Garching, Germany
    \and
      Universit\`a di Bologna, Dipartimento di Fisica e Astronomia,
      via Berti Pichat 6/2, 40127 Bologna, Italy
}

   \date{Received: 30/10/2014; accepted 30/1/2015}

  \abstract{
    Wide area X-ray and far infrared surveys are a fundamental tool to
    investigate the link between AGN growth and star formation,
    especially in the low-redshift universe ($z\lesssim 1$). The
    \textit{Herschel} Terahertz Large Area survey (H-ATLAS) has
    covered 550 deg$^2$ in five far-infrared and sub-mm bands, 16
    deg$^2$ of which have been presented in the Science Demonstration
    Phase (SDP) catalogue.
    Here we introduce the \xmm\ observations in H-ATLAS SDP area,
    covering 7.1 deg$^2$ with flux limits of $2\e{-15}$, $6\e{-15}$
    and $9\e{-15}$ \ergscmq\ in the 0.5--2, 0.5--8 and 2--8 keV bands,
    respectively.
    We present the source detection and the catalogue, which includes
    1700, 1582 and 814 sources detected by \emld\ in the 0.5--8,
    0.5--2 and 2--8 keV bands, respectively; the number of unique
    sources is 1816. We extract spectra and
    derive fluxes from power-law fits for 398 sources with more than
    40 counts in the 0.5--8 keV band. We compare the best-fit fluxes
    with the catalogue ones, obtained by assuming a common photon
    index of $\Gamma=1.7$; we find no bulk difference between the
    fluxes, and a moderate dispersion of $s=0.33$ dex.  Using wherever
    possible the fluxes from the spectral fits, we derive the 2--10
    keV \lognlogs, which is consistent with a Euclidean
    distribution. Finally, we release computer code for the tools
    developed for this project.
  }


     \authorrunning{P. Ranalli et al.}
     \titlerunning{The XMM-ATLAS survey}

   \maketitle

\begin{table*}
\centering
\caption[]{XMM-ATLAS observations. The columns show: observation ID;
  date; total exposure times (ks) for the MOS1, MOS2 and PN cameras;
  exposure times (ks) after high background filtering for the MOS1,
  MOS2 and PN cameras; astrometry corrections applied (arcsec).}
\label{tab:obsid}
\begin{tabular}{ccrrrrrrrr}
Obsid      &Date       &MOS1  &MOS2  &PN  &MOS1 clean &MOS2 clean &PN clean &$\Delta$RA & $\Delta$Dec\\
\hline
0725290101 &2013-05-05 &113   &113   &110 &93         &96         &71       &---        &---    \\
0725300101 &2013-05-07 &113   &113   &110 &101        &100        &61       &-0.44      &-0.47  \\
0725310101 &2013-05-21 &110   &110   &112 &99         &100        &94       &-1.43      &+0.41  \\
\end{tabular}
\end{table*}

\section{Introduction}
\label{sec:intro}

Over more than one decade, there has been growing evidence for a
coeval growth of galaxies and their central black holes
\citep[see review by][]{alexander2012}. A tight correlation between the masses of the
black hole and the galaxy bulge has been found
\citep[e.g.][]{ferrares2000,gebhardt2000,zubovas2012}. Theoretical
models suggest that feedback processes are at work to set the link
\citep{dimatteo2005,hopkins2006}; some models also suggest a
fundamental role of mergers in setting up AGN
\citep{hopkins2008a}. Nuclear obscuration and intense star formation
may characterize the initial phases of AGN activity
\citep{silk1998,menci2008,lamastra2013}.

AGN growth seems to happen in two major modes: the radiative and the
kinetic mode, the first operating close to the Eddington limit and
with high radiation efficiency, the latter at lower rates
\citep[see review by][]{fabian2012}.  Similarly, galaxies build their stellar mass
either through starburst episodes (with star formation happening on
short timescales), or through secular star formation. The growth of
both AGN and galaxy should ultimately be driven by the supply of cold gas
\citep{kauffmann2009,mullaney2012b}.

The fraction of galaxies hosting an AGN increases with far-infrared
luminosity or star formation rate (SFR)
\citep{kim1998,veilleux1999,tran2001}, reaching the 50--80\% among
Luminous InfraRed Galaxies (LIRG) and UltraLuminous InfraRed Galaxies
(ULIRG)
\citep{alexander2008,lehmer2010,nardini10,nardini2011,alonso-herrero2012,ruiz2013};
IR and X-ray observations are the key for the identification of AGN.

It has been suggested that the specific SFR (i.e., the SFR divided by
the stellar mass of the galaxy) may also be involved in the AGN
growth/star formation link, because a correlation was observed
between AGN luminosity and specific SFR \citep{lutz2010}; however the
latter correlation seems to hold only for the most active systems at
redshift $z\gtrsim 1$ \citep{mullaney2012a,rovilos2012}.  Similar uncertainties
shroud a possible correlation between AGN luminosity and nuclear
obscuration
\citep{georgakakis2006,rovilos2007,trichas2009,rovilos2012} and the
effectiveness of colour-magnitude diagrams to inspect the evolutionary
status of the host galaxies of AGN
\citep{brusa2009,cardamone2010b,pierce2010}.

The situation in the local universe is mostly unclear, as the
aforementioned studies were based on deep, pencil-beam surveys.
Wide area surveys are hence needed to probe larger volumes of the
low-redshift universe, and build sizable samples of rarer objects. 

The \textit{Herschel} Terahertz Large Area survey (H-ATLAS) is the largest Open
Time Key Project carried out with the \textit{Herschel} Space Observatory
\citep{eales2010-hatlas}, covering $\sim 550$ deg$^2$ with both the
SPIRE and PACS instruments in five far-infrared and sub-mm bands (100,
160, 250, 350 and 500 $\mu$m). The 250 $\mu$m source catalogue of the
Science Demonstration Phase (SDP) is presented in
\citet{rigby2011-hatlas-sdp}, covering a contiguous area of $\sim 16$ deg$^2$
which lies within one of the regions observed by the Galaxy And Mass
Assembly (GAMA) survey \citep{driver2009-gama,baldry2010}.

In this paper, we present XMM-ATLAS i.e.\ the \xmm\ observations and
source catalogue in the H-ATLAS area. \xmm\ observed 7.1 deg$^2$
within the H-ATLAS SDP area, making the XMM-ATLAS one of the largest
contiguous areas of the sky with both \xmm\ and \textit{Herschel}
coverage.

The only other wide X-ray survey with size and
  \textit{Herschel} coverage comparable to XMM-ATLAS is the 11 deg$^2$
  XMM-LSS \citep{pierre2004lss,chiappetti2013} which was observed with
  SPIRE and PACS by the HerMES project \citep{oliver2012hermes}. The
  Stripe-82 survey covered 10.5 deg$^2$ with \xmm\footnote{Reaching
    16.5 deg$^2$ when considering both \xmm\ and \chandra\ data.}
  \citep{lamassa2013}, though the pointings are not contiguous and it
  only has SPIRE coverage \citep{viero2014}. The 2 deg$^2$ COSMOS
  survey, observed by both \xmm\ \citep{xmm-cosmos} and
  \chandra\ \citep{ccosmos-cat,civano2013head}, was also covered by HerMES with
  SPIRE only \citep{oliver2012hermes}.

For the bright sources in the XMM-ATLAS catalogue, we
extract spectra and derive fluxes from the spectral fits. The latter
fluxes are used to build the \lognlogs, thus removing as much as
possible any bias which might come from using a single
count-rate-to-flux conversion factor as done for the catalogue.

In Sect.~\ref{sec:observations}, we describe the observations and
data reduction; in Sect.~\ref{sec:detection}, we illustrate the source
detection; in Sect.~\ref{sec:catalogue} we present the source
catalogue in the 0.5--2, 0.5--8 and 2--8 keV bands; in
Sect.~\ref{sec:spectra} we derive spectra for the bright sources
and compare the fluxes from the spectral fits with those from the
catalogue; in Sect.~\ref{sec:lognlogs} we compute the
\lognlogs\ using, where available, the fluxes from the spectral
fits. Finally, in Sect.~\ref{sec:conclusion} we present our
conclusions.

\section{Observations and data reduction}
\label{sec:observations}

The XMM-ATLAS field is centred at 9h~4m~30.0s +0d~34m~0s, and covers
7.101 deg$^2$, with a total exposure time of 336 ks (in the MOS1
camera). The observations were performed in mosaic mode,
i.e.\ shifting the pointing by $15 \arcmin$ every $\sim 10$ ks. A
total of 93 pointings were done, divided in 3 obsids of 31 pointings
each.

The SAS (version 13.0) tools \texttt{emproc} and
  \texttt{epproc} were used to produce a single event file per obsid
  per camera. Such an event file can be directly used to obtain images
  and exposure maps of the part of the mosaic covered in the
  obsid. However, the file needs to splitted in order to produce
  images and exposure maps of individual pointings (see
  Sect.~\ref{sec:splitmosaic}).  

We extracted lightcurves in the 10--13 keV interval with a
  100~s bin size  to check for
high-background periods, which were identified and removed by doing a
$3\sigma$-clipping of the lightcurve, as done in \citet[hereafter
  R13]{cdfscat} for the XMM-CDFS. After cleaning, the total
exposure is 293 ks (MOS1).

\begin{figure}
  \centering
  \includegraphics[width=\columnwidth]{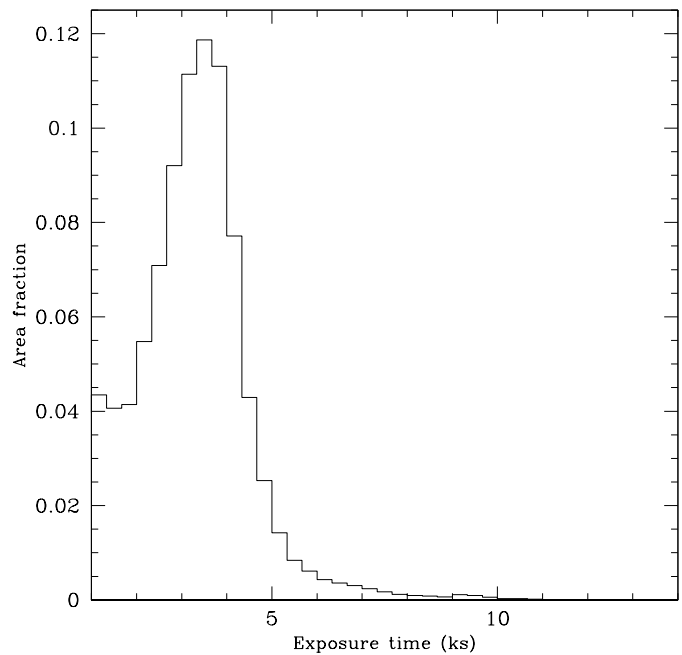}
  \caption{ Distribution of the exposure times in the XMM-ATLAS mosaic
    (bin size: 1 ks). The times are the average between MOS1,
    MOS2 and PN, for the 0.5--8 keV band; the vertical axis shows the
    fraction of pixels with a given exposure.  }
  \label{fig:exphist}
\end{figure}

The obsid, dates and exposure times before and after the
high-background cleaning, are listed in Table~\ref{tab:obsid}; note
that these times refer to the mosaic-mode obsid.  The average exposure
time for any location in the final mosaic is 3.0 ks (in the 0.5--8 keV
band and after the high-background cleaning); the maximum exposure is
11 ks. A histogram of the exposure times is shown in
Fig.~\ref{fig:exphist}.

\begin{figure*}
  \centering
  \includegraphics[width=.49\textwidth]{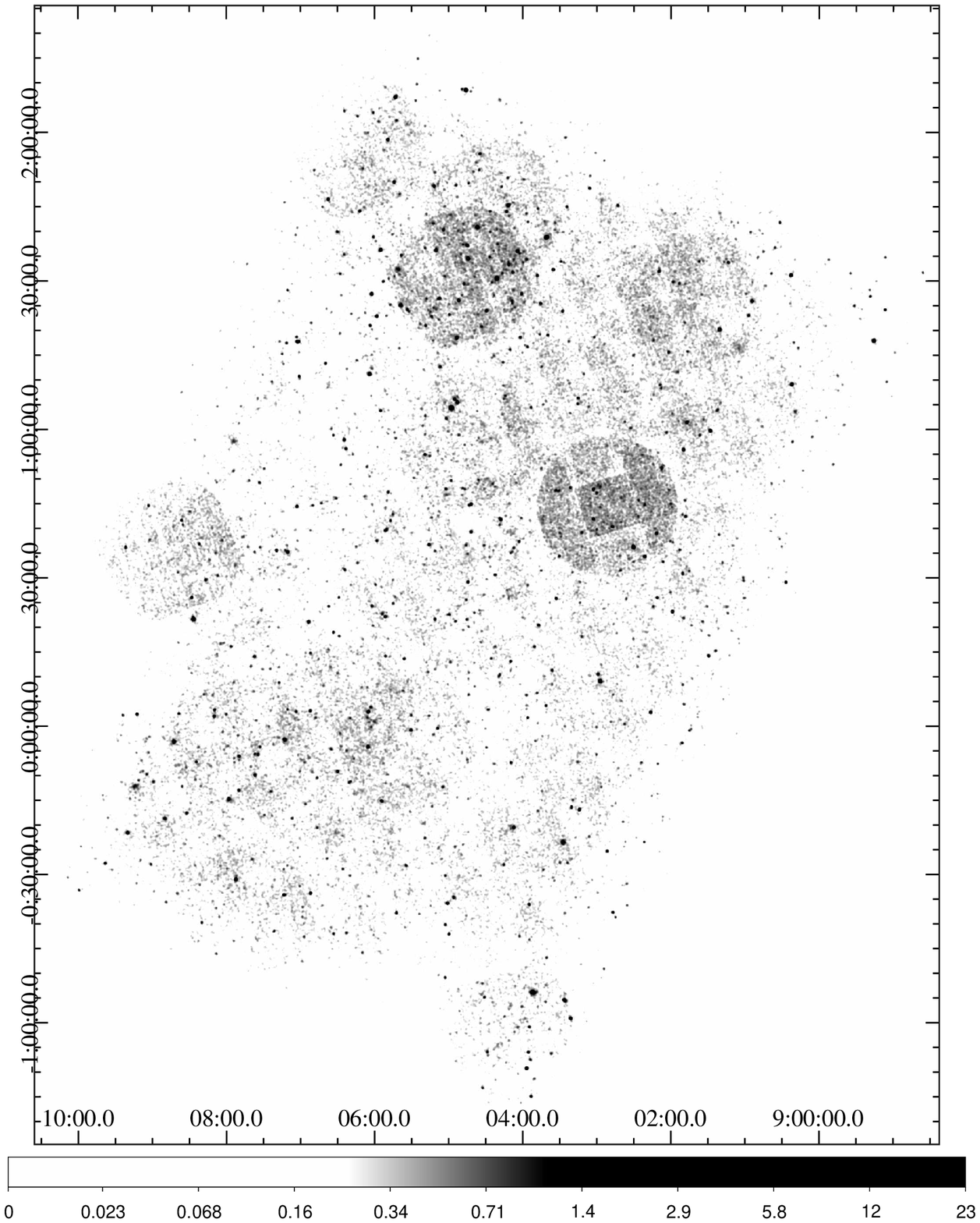}
  \includegraphics[width=.49\textwidth]{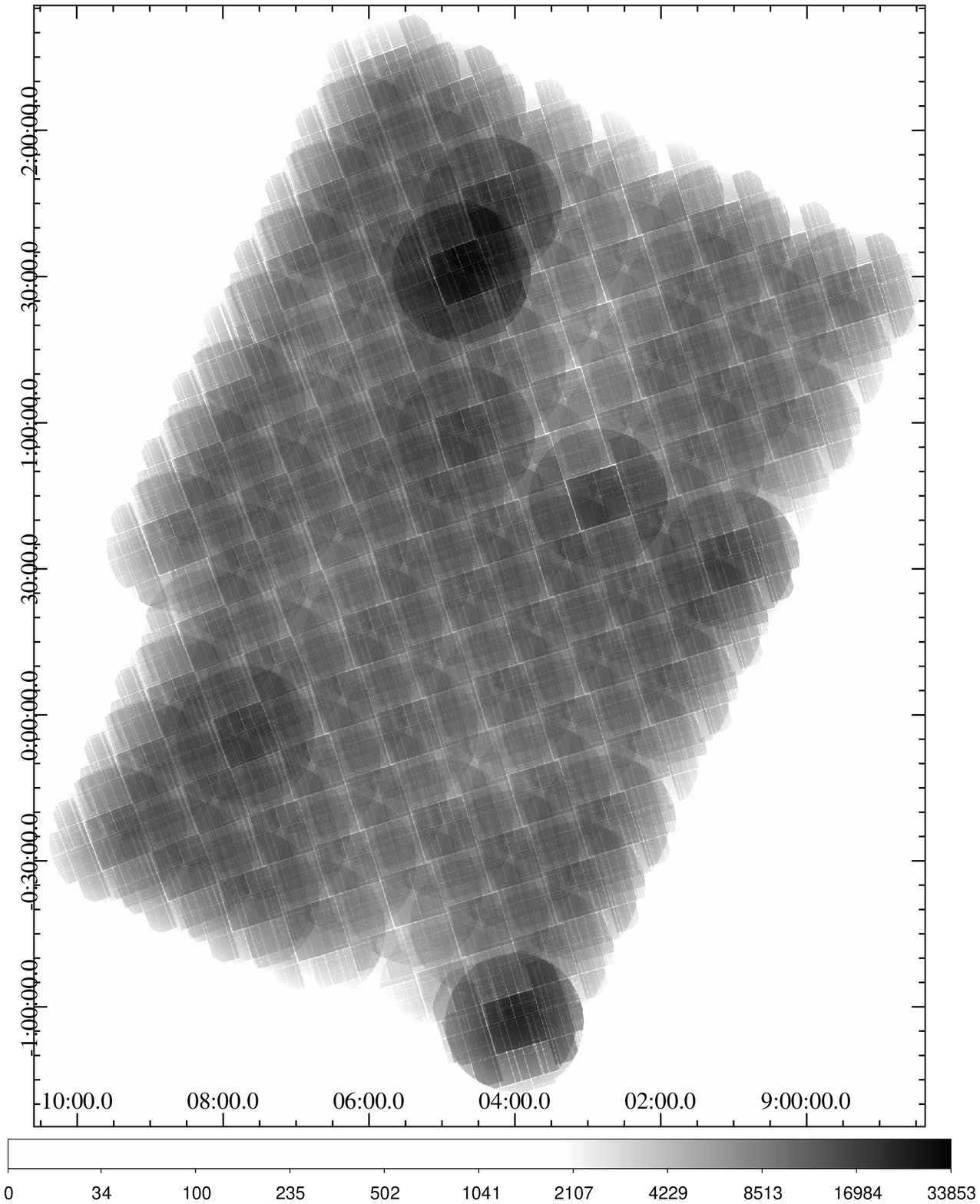}
  \caption{ \textit{Left:} image of the entire mosaic of the XMM-ATLAS
    observations in the 0.5--8 keV band. The greyscale wedge shows the
    photon counts (truncated; the actual maximum is 117 counts). A
    Gaussian smoothing with $20\arcsec$ radius has been applied to
    enhance the sources. Some areas with longer exposure times and/or
    higher background are visible.  \textit{Right:} exposure map in
    the 0.5--8 keV band. The greyscale wedge (maximum value: 33.9 ks)
    shows the sum of the MOS1, MOS2 and PN exposure times in s,
    corrected for vignetting and bad pixels.  }
  \label{fig:mosaic}
\end{figure*}

To check if astrometric corrections were needed, we performed an
initial source detection run with the SAS \ewav\ program (with
  a $5\sigma$ threshold) on the PN data of each of the three obsids,
and cross-correlated the resulting lists with the sample of QSO from
the Sloan Digital Sky Survey \citep[SDSS DR7;][]{schneider2010}.
The search radius was $15\arcsec$, and most matches were found
  within $5\arcsec$.
We found that obsid 0725290101 (for
    which 26 matches were found) needed no correction; obsid
  0725300101 (35 matches) had a shift of $0.58\arcsec$ and
  obsid 0725310101 (39 matches) of $1.14\arcsec$. Event and
  attitude files were corrected (see Table\ref{tab:obsid}) before
  proceeding further.

Images for the MOS1, MOS2 and PN cameras were accumulated for
  each obsid and for each of the 0.5--2, 2--8 and 0.5--8 keV bands
with a pixel size of $4\arcsec$ and summed together. Some energy
intervals corresponding to known instrumental spectral lines
\citep[][see also R13] {kuntz2008} were excluded: 1.39--1.55 keV (Al);
1.69--1.80 keV (Si, MOS only); 7.35--7.60 and 7.84--8.28 keV (Cu
complex\footnote{The same complex also includes a line in the
  8.54--9.00 keV interval, which is outside of the bands considered
  here.}, PN only).  Images of the entire mosaic were
obtained by summing the images and exposure maps from the three
obsids; those relative to the 0.5--8 keV band are shown in
  Fig.~\ref{fig:mosaic}.

\section{Source detection}
\label{sec:detection}

Sources were detected with a two-stage process, with a first
  pass at low significance to get a list of candidate detections, and
  a second pass to rise the significance threshold and derive accurate
  source parameters. Between the two passes, and because the second
  pass needs an input catalogue, we identify the sources detected in
  more than one band.  The detection method is based on the one used
  by \cite{xmm-cosmos} and R13, which is a variant on the standard
  \xmm\ detection procedure, adapted to the much larger area of
  XMM-ATLAS.

\subsection{First detection pass}

In the first pass, the SAS wavelet detection program
\ewav\ was run separately on each of the 0.5--2, 2--8 and 0.5--8 keV
images of the entire mosaic, with a significance threshold of $4\sigma$
and the default wavelet scales (minimum 2 pixels, maximum 8 pixels,
with a pixel size of $4\arcsec$).

\subsection{Matching sources detected in more than one band}
\label{sec:LRmatch}

Sources detected in more than one band were identified with
  the likelihood ratio (LR) technique. Given a distance $r$ between
  two candidate counterparts, normalised by the uncertainty on the
  position, LR$(r)$ is defined as the ratio between the probability of
  having a real counterpart at $r$ over the probability of having a
  spurious counterpart at $r$
  \citep{pineau2011}\footnote{\citet{pineau2011} also developed an
    Aladin plugin, which we used for this paper, available at
    \texttt{saada.u-strasbg.fr/docs/fxp/plugin/}}.  Similarly, the
  association reliability is defined as the probability
  $P(H_\mathrm{true}|r)$ of the association being true ($H_\mathrm{true}$)
  conditioned on $r$ \citep[the relationship between LR and
    reliability is explicited in][eq.~11 and Appendix C]{pineau2011}.

  While the LR is
  usually applied to search for counterparts in independent bands
  (e.g., optical counterparts of X-ray sources), here the bands are
  not independent (we compare the 0.5--8 keV list to the 2--8 keV one;
  and the 0.5--8 keV list to the 0.5--2 keV one). In this setting,
  and taking the latter case as an example, we are testing if the
  position of the source in both bands is close enough (within errors)
  to declare that they are the same source, against the possibility
  that the source of the additional 2--8 keV photons is significantly
  different from that of the 0.5--2 keV photons.  In a shallow survey
  such as XMM-ATLAS,

  the main advantage in using the LR method over a
  nearest-neighbour match is that LR gives a likelihood for the match,
  which may help in follow-up studies, e.g.\ when searching for optical
  counterparts.

The match candidates were selected on the basis of their positional
errors; the probability of association (items \ref{item:assocproba1}
and \ref{item:assocproba2} in the list in Sect.~\ref{sec:catalogue})
was calibrated by estimating a spurious LR histogram.
Source IDs were assigned at this stage; matched sources were
  assigned a single ID and considered as a single sources from here onwards.

\subsection{Second detection pass, general settings}

In the second pass, we used the SAS \emld\ program to validate the
detection, refine the coordinates and obtain maximum-likelihood
estimates of the source counts, count rates and fluxes.

\emld\ is at its core a PSF fitting code. Originally developed
  for ROSAT \citep{cruddace1988,hasinger1993}, its current
  version\footnote{The XMM SAS reference manual for \emld;
    \texttt{xmm.esac.esa.int/sas/13.0.0/doc/emldetect/node3.html}} has
  been improved and optimized for \xmm. In particular, we mention the
  ability to operate on several individual, overlapping pointings at
  the same time, and properly account for variations in PSF shape and
  in vignetting (as the telescope is moved between pointings, a source
  may be present in more than one pointing, but may fall on different
  detector coordinates). Rather than working on a single image
  containing the entire mosaic (like \ewav), \emld\ runs on images of
  individual pointings (see Sect.~\ref{sec:splitmosaic}).

Although \emld\ can run on different energy bands at the same
  time, we did not use this feature, because of the large number of
  pointings and of \emld\ limitations (see below), but rather we ran
  it three times, one per each band. The bands are the same as for
  \ewav. In all runs, we used the same input list of sources, made of
  all matched sources and all unmatched sources, regardless of the
  bands in which they were detected in the first pass.

The \emld\ minimum likelihood was set at $L=4.6$, as in R13, which
corresponds to a false-detection probability of $1.01\e{-2}$. Together
with the $4\sigma$ threshold for \ewav, for the final catalogue this
yields a joint significance between $4\sigma$ and $5\sigma$, but which
cannot be further constrained without simulations (see discussion in
R13, where a joint significance of $4.8\sigma$ was estimated; however
this number should not be immediately applied to XMM-ATLAS because of
the different statistical properties of the background, due to the
different depth of this survey vs.\ the XMM-CDFS).

The count rate to flux conversion factors were derived assuming a
power-law spectrum with photon index $\Gamma=1.7$ and Galactic
absorption of $N_H=2.3\e{20}$ cm$^{-2}$, and by weighting the
responses of the MOS1, MOS2 and PN cameras; they are $5.83\e{-12}$,
$3.25\e{-12}$ and $1.17\e{-11}$ erg~cm$^{-2}$ for the 0.5--8, 0.5--2
and 2--8 keV bands, respectively.

The 8 keV upper threshold was chosen to improve the signal/noise ratio
of detections.
While quantities
such as counts, rates and fluxes are quoted in the catalogue for the
0.5--8 and 2--8 keV bands, fluxes in the 0.5--10 and 2--10 keV bands
can be immediately obtained by assuming the same $\Gamma=1.7$ model
spectrum, which yields the following conversion factors for fluxes
$F$: $F(0.5-10)/F(0.5-8)=1.13$ and $F(2-10)/F(2-8)=1.20$.

All \emld\ parameters, except for what mentioned above
  (coordinate fitting\footnote{Using the parameter
      \texttt{fitposition=yes}.}, likelihood threshold, energy bands
  and conversion factors) were left at their default values. The
  inputs to \emld\ are images, exposure maps (see
  Sect.~\ref{sec:splitmosaic}), background maps (see
  Sect.~\ref{sec:bkgmaps}) and the list of candidate sources from
  Sect.~\ref{sec:LRmatch}.

\begin{figure}
  \centering
  \includegraphics[height=\columnwidth,angle=-90,bb=58 98 529 592,clip]{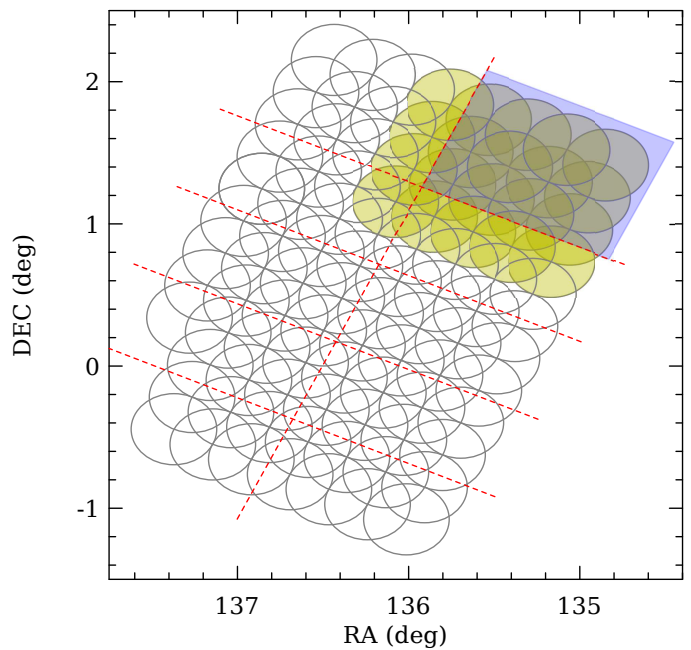}
  \caption{ Detection grid for \emld. The grey circles have a radius
    of $15\arcmin$ and show the FOV of the individual pointings. The
    dashed red lines show the grid. The cyan rectangle highlights one
    of the detection cells. The filled yellow circles are the
    pointings which intersect the cell, and which are used for the
    detection. Only the sources whose coordinates fall into the cell
    (the rectangle) are kept. This process is repeated for all cells
    and the final catalogue is the union of the sources from all
    cells.  }
  \label{fig:detectiongrid}
\end{figure}

\subsection{Splitting a mosaic in individual pointings}
\label{sec:splitmosaic}
 
To use \emld, we had to split each mosaic-mode obsid event
  file into the individual pointings; for this purpose we used the
SAS tool \texttt{emosaic\_prep}. This tool did not set the RA\_PNT and
DEC\_PNT keywords in the individual pointings, which were needed for
further processing; however, the coordinates were obtained by
inspecting the attitude files.

Images and exposure maps were extracted from each pointing's
  event file in the same bands used for \ewav.  The SAS task
  \texttt{eexpmap} was used to compute the exposure maps; they are in
  units of s, and include corrections for vignetting and bad
  pixels. Non-vignetted exposure maps were also computed, to be used
  when producing background maps (Sect.~\ref{sec:bkgmaps}).

 Since the PSF shapes and vignetting are very similar among
  the three \xmm\ EPIC cameras, for each pointing we summed together
the images from MOS1, MOS2 and PN; then we did the same for the
exposure and background maps (see below).

\subsection{Background maps}
\label{sec:bkgmaps}

Background maps for the individual pointings were obtained with the
same method used by the XMM-COSMOS team
\citep{xmm-cosmos}. All input sources were excised from the
  pointing images, producing so-called ``cheese images'', in a number
  of one per pointing per energy band. A model made of a flat
  (i.e. non-vignetted, but including chip gaps and dead pixels) and a
  vignetted component (representing the particle and cosmic
  backgrounds, respectively) was fit to the cheese images to obtain
  the background maps.

\subsection{Running \emld\ on a wide mosaic}
\label{sec:whygriddetect}

\emld\ can detect sources on a limited number of overlapping
pointings; however, the number of individual pointings in XMM-ATLAS is
too large to be analysed together.  Therefore, we divided the ATLAS
field in a grid. For each cell in the grid (for example, the rectangle
in Fig.~\ref{fig:detectiongrid}), we identified the pointings whose
FOV intersected the cell (filled circles in
Fig.~\ref{fig:detectiongrid}), and ran \emld\ on them together. In
this way, the number of pointings was within the limit allowed by
\emld. From the resulting detection list, we only kept the sources
whose coordinates fell into the cell. We checked that no source
appears twice, or was lost, by searching for source pairs within
$3\arcsec$ from the cracks. We repeated this procedure for each of the
0.5--8, 0.5--2, and 2--8 keV bands. The final catalogue is the union
of the detections from all cells. 
This approach is similar to the one
adopted by \citet{lamassa2013}.

The programme which runs \emld\ in the cells, {\tt griddetect}, is
available (see Appendix~\ref{sec:software}).

\subsection{Inspection of close groups}
\label{sec:inspection}

Very close groups of sources with separation much shorter than
  the PSF width may arise in particular conditions because \emld\ is
  fitting the source positions and the detection runs are done
  separately for the three energy bands. For example, let A and B be
  two sources from the input catalogue which are present in the same
  \emld\ detection cell\footnote{The detection cell size is $10\times
    10$ pixels$=40\arcsec\times 40\arcsec$.}. If at the input location
  of A there are not enough counts for a significant detection by
  \emld, the programme may try and fit the coordinates until
  incorrectly assigning them to B's position; this will then prevent B
  from getting a detection at the same position and in the same
  band. This might happen in one energy band but not in another,
  depending on the details of the two sources (e.g., input
  coordinates, location of the photons, hardness ratio).  Therefore,
  the same \emld\ coordinates would be shared by sources A and B,
  while their \ewav\ coordinates would still be different.  

In order to screen and correct for this effect, we have looked
  for close groups of sources whose \emld\ coordinates were closer
  than their $5\sigma$ error. We identified 27 groups containing 47
  sources in total, and visually inspected all of them. For 22 groups, we
  shuffled the source ids to have the \emld\ positions match the
  \ewav\ ones. Five groups looked like genuine different sources in
  crowded areas.

\subsection{Systematic error on astrometry}
\label{sec:eml-astrometry}

To find whether there is any residual systematic error in the
XMM-ATLAS astrometry, we cross-correlated the final catalogue with the
SDSS QSOs \cite[DR7;][]{schneider2010}. We found a bulk shift of
$0.83\arcsec$ in RA and $-0.29\arcsec$ in DEC (XMM-ATLAS coordinate
minus SDSS coordinate; root mean square deviation$\,= 2.1\arcsec$). We subtracted the above
values from the XMM-ATLAS coordinates (``RA'' and ``DEC'' columns
only, see next Sect.), which should therefore not present any further
shift.

\begin{figure}
  \centering
  \includegraphics[width=\columnwidth,bb=25 158 414 553,clip]{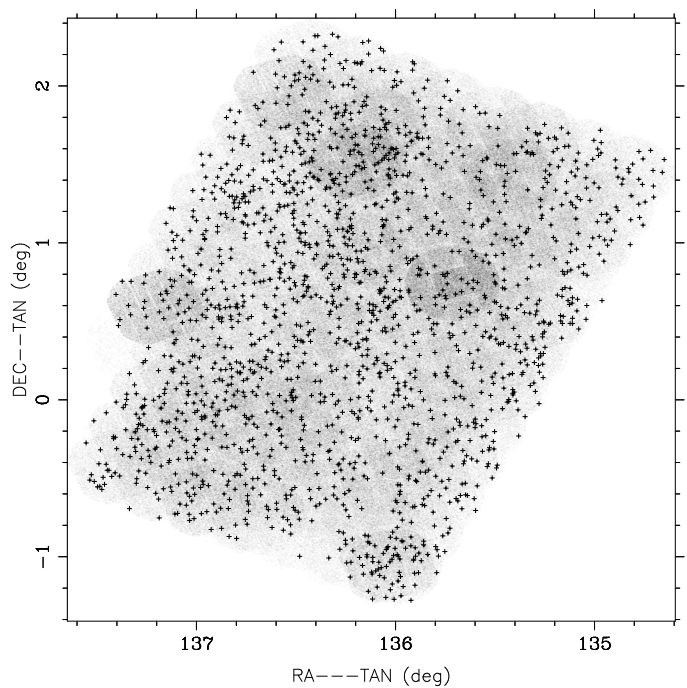}
  \caption{ Positions of detected sources, superimposed on the 0.5--8
    keV image.   }
  \label{fig:img-w-src}
\end{figure}

\section{X-ray catalogue}
\label{sec:catalogue}

The XMM-ATLAS catalogue includes 1700, 1582 and 814 sources
detected by \emld\ in the 0.5--8, 0.5--2 and 2--8 keV bands,
respectively. The number of unique sources is 1816.  The flux
limits, defined as the flux of the faintest detected sources, are
$2\e{-15}$, $6\e{-15}$ and $9\e{-15}$ \ergscmq\ in the 0.5--2, 0.5--8
and 2--8 keV bands, respectively. The positions of the detected
sources are shown in Fig.~\ref{fig:img-w-src}, superimposed on a
0.5--8 keV image of the mosaic.

In addition, we list a number of 175, 103 and 47 sources detected in
the above bands by \ewav\ but not confirmed by \emld; in the following
we refer to them as supplementary sources. The number of unique
supplementary sources is 234.

All coordinates are in the J2000 reference system. The RA and
  DEC columns are registered to the reference frame of the SDSS-DR7 QSOs
  \citep{schneider2010}. All other coordinate columns are not
  registered and present a shift;
  see Sect.~\ref{sec:eml-astrometry}.

The catalogues are available in electronic form from the Centre de
Donne\'es Astronomiques de Strasbourg (CDS), and from the XMM-ATLAS
website%
\footnote{\texttt{xraygroup.astro.noa.gr/atlas.html}}, where we will publish updates should
they become available.

The column description is as follows; the symbols \ML\ and \W\ mark
whether the column was derived from running \emld\ or \ewav,
respectively.

\begin{enumerate}
\setlength{\itemindent}{4ex}
\item[1.] IAU\_IDENTIFIER --- source identifier following International
  Astronomical Union conventions;
\item[2.] ID --- unique source number;
\item[3--5.] ID058, ID052, ID28
\setcounter{enumi}{5}
\item RA --- \ML\ right ascension (degrees) from the 0.5--8 keV band,
  if available; else, from the 0.5--2 keV band, if available; else,
  from the 2--8 keV band. This column has been corrected for
  astrometry and registered to the SDSS-DR7 QSO framework;
\item DEC --- \ML\ declination (degrees) as above. This column has been corrected for
  astrometry and registered to the SDSS-DR7 QSO framework;
\item RADEC\_ERR --- \ML\ error on position\footnote{This column was
  obtained by dividing the \emld\ RADEC\_ERR by $\sqrt{2}$.  The
  RADEC\_ERR from \emld\ is computed as
  $\left(\mathrm{RA\_ERR}^2+\mathrm{DEC\_ERR}^2\right)^{1/2}$. However,
  when two normalised one-dimensional gaussians of sigma $s$ are
  combined, the corresponding normalised bi-dimensional gaussian also
  has sigma $s$, not $\sqrt{2}s$, which is what one would get with the
  expression above. Therefore, we divided the \emld\ value by
  $\sqrt{2}$.}  (arcsec; $1\sigma$);
\item WAV\_RA    --- \W\  merged right ascension (degrees) from the
  likelihood ratio;
\item WAV\_DEC   --- \W\ merged declination (degrees);
\item WAV\_RADEC\_ERR --- \W\ merged error on position (arcsec; $1\sigma$)
\item WAV\_RA058   --- \W\ right ascension (degrees) in the 0.5--8 keV band;
\item WAV\_DEC058  --- \W\ declination (degrees) in the 0.5--8 keV band;
\item WAV\_RA052   --- \W\ right ascension (degrees) in the 0.5--2 keV band;
\item WAV\_DEC052  --- \W\ declination (degrees) in the 0.5--2 keV band;
\item WAV\_RA28    --- \W\ right ascension (degrees) in the 2--8 keV band;
\item WAV\_DEC28   --- \W\ declination (degrees) in the 2--8 keV band;      
\item RA058        --- \ML\ right ascension (degrees) in the 0.5--8 keV band;        
\item DEC058       --- \ML\ declination (degrees) in the 0.5--8 keV band;
\item RA052        --- \ML\ right ascension (degrees) in the 0.5--2 keV band;         
\item DEC052       --- \ML\ declination (degrees) in the 0.5--2 keV band;             
\item RA28         --- \ML\ right ascension (degrees) in the 2--8 keV band;           
\item DEC28        --- \ML\ declination (degrees) in the 2--8 keV band;               
\item ASSOC\_RELIAB058052  --- association reliability between the
  (WAV\_RA058, WAV\_DEC058) and   (WAV\_RA052, WAV\_DEC052)
  coordinates;
  \label{item:assocproba1}
\item ASSOC\_RELIAB05828  --- association reliability between the
  (WAV\_RA058, WAV\_DEC058) and   (WAV\_RA28, WAV\_DEC28) coordinates;
  \label{item:assocproba2}
\item[26--28.] SCTS058, SCTS052, SCTS28      --- \ML\ sum of the net
  source counts from MOS1+MOS2+PN in the 0.5--8, 0.5--2 and 2--8 keV
  bands, respectively;
\item[29--31.] SCTS\_ERR058, SCTS\_ERR052, SCTS\_ERR28 --- \ML\ errors
  on SCTS058, SCTS052, SCTS28 ($1\sigma$);
\item[32--34.] RATE058, RATE052, RATE28      --- \ML\ net count rates in
  the 0.5--8, 0.5--2, 2--8 keV bands, averaged over the three cameras;
\item[35--37.] EXP\_MAP058, EXP\_MAP052, EXP\_MAP28  --- exposure times in the 0.5--8, 0.5--2, 2--8 keV bands, summed over the three cameras;
\item[38--40.] BG\_MAP058, BG\_MAP052, BG\_MAP28   ---
  \ML\  background counts/arcsec$^2$ in the 0.5--8, 0.5--2, 2--8 keV
  bands, summed over the three cameras;         
\setcounter{enumi}{40}
\item[41--43.] FLUX058, FLUX052, FLUX28  --- \ML\ flux in the 0.5--8,
  0.5--2, 2--8 keV bands (\ergscmq);         
\item[44--46.] FLUX\_ERR058, FLUX\_ERR052, FLUX\_ERR28 --- \ML\ error
  on FLUX058, FLUX052, FLUX28 ($1\sigma$);        
\item[47--49.] DETML058, DETML052, DETML28    --- \ML\ detection likelihoods in the 0.5--8, 0.5--2, 2--8 keV bands;      

\item[50--51.] EXT058, EXT052 -- \ML\ source extent in the 0.5--8 and
  0.5--2 keV bands\footnote{EXT28 and related columns are not included since no source was found to be extended in the 2--8 keV band.} ($\sigma$ of Gaussian model in pixels; 1 pixel = $4\arcsec$);
\item[52--53.] EXT\_ERR058, EXT\_ERR052 -- \ML\ error on EXT058,
  EXT052 ($1\sigma$);
\item[54--55.] EXT\_ML058, EXT\_ML052 -- \ML\ likelihood of extent in
  the 0.5--8 and 0.5--2 keV bands;

\setcounter{enumi}{55}
\item HR           --- hardness ratio, computed from $S=\mathrm{SCTS052}$ and $H=\mathrm{SCTS28}$ as $\mathrm{HR}=(H-S)/(H+S)$;
\item HR\_ERR      --- error on HR  ($1\sigma$; see below);  \label{item:HR_ERR}
\end{enumerate}

The columns from 58 to 87 contain source properties (counts, count
rates, fluxes, exposure times, background, wavelet detection scale,
source extent) from \ewav; while we report
them for all sources, they are actually interesting only for the
supplementary sources.

\begin{enumerate}
\setlength{\itemindent}{4ex}
\item[58--60.] WAV\_SCTS058, WAV\_SCTS052, WAV\_SCTS28 --- \W\ sum of the net
  source counts from MOS1+MOS2+PN in the 0.5--8, 0.5--2 and 2--8 keV
  bands, respectively;
\item[61--63.] WAV\_SCTS\_ERR058, WAV\_SCTS\_ERR052, WAV\_SCTS\_ERR28
  --- \W\ errors on SCTS058, SCTS052, SCTS28 ($1\sigma$);
\item[64--66.] WAV\_RATE058, WAV\_RATE052, WAV\_RATE28 --- \W\ net count rates in
  the 0.5--8, 0.5--2, 2--8 keV bands, averaged over the three cameras;
\item[67--69.] WAV\_EXP\_MAP058, WAV\_EXP\_MAP052, WAV\_EXP\_MAP28 ---
  \W\ exposure times in the 0.5--8, 0.5--2, 2--8 keV bands, summed over the three cameras;
\item[70--72.] WAV\_BG\_MAP058, WAV\_BG\_MAP052, WAV\_BG\_MAP28 ---
  \W\ background counts/arcsec$^2$ in the 0.5--8, 0.5--2, 2--8 keV
  bands, summed over the three cameras;
\item[73--75.] WAV\_FLUX058,  WAV\_FLUX052,  WAV\_FLUX28 ---
  \W\ fluxes in the 0.5--8, 0.5--2, 2--8 keV bands (\ergscmq);
\item[76--78.] WAV\_FLUX\_ERR058, WAV\_FLUX\_ERR052, WAV\_FLUX\_ERR28
  --- \W\ errors on WAV\_FLUX058, WAV\_FLUX052, WAV\_FLUX28 ($1\sigma$);
\item[79--81.] WAV\_WSCALE058,  WAV\_WSCALE052,  WAV\_WSCALE28 ---
  \W\  wavelet detection scale (pixels; 1 pixel = $4\arcsec$);
\item[82--84.] WAV\_EXTENT058, WAV\_EXTENT052,  WAV\_EXTENT28 ---
  \W\ source extent (pixels);
\item[85--87.] WAV\_EXT\_ERR058, WAV\_EXT\_ERR052, WAV\_EXT\_ERR28 ---
  \W\ errors on WAV\_EXTENT058, WAV\_EXTENT052,  WAV\_EXTENT28
  ($1\sigma$; pixels).
\end{enumerate}

\smallskip

The error on the hardness ratio (column \ref{item:HR_ERR}, HR\_ERR) is
defined as 
\begin{equation}
\mathrm{HR\_ERR}=
2 \frac{
  \sqrt{ (H \, \sigma_\mathrm{S})^2 + (S \, \sigma_\mathrm{H})^2}
}{
  (H+S)^2
}
\end{equation}
where $H$ and $S$ are the hard (SCTS28) and soft net counts (SCTS052),
respectively, and $\sigma_\mathrm{H}$ (SCTS\_ERR28) and
$\sigma_\mathrm{S}$ (SCTS\_ERR052) are their errors.

\section{Spectra}
\label{sec:spectra}

To investigate the effects of the assumption of a single model
spectrum (power-law with $\Gamma=1.7$, see Sect.~\ref{sec:detection}),
we extracted spectra for 555 sources with more than 40 counts in
  the 0.5--8 keV band\footnote{The number of sources with
    SCTS058$\ge40$ is 569, but 14 were dropped because they are in
    crowded areas and their spectra would include contributions from
    neighbour sources.}.

The extraction regions are circles, whose radii were chosen by
maximizing the signal/noise ratio for each source. The background
regions are annuli, of inner and outer radii equal to 1.5 and 2 times
the circle radius, respectively. In the case of close sources, the
overlapping area was excised from the source and background regions.
The programme which defines the regions, {\tt autoregions}, is
available (see Appendix~\ref{sec:software}).

Spectra were extracted using the {\tt cdfs-extract} tool (see Appendix
\ref{sec:software}), which takes care of identifying the available
combinations of event files and source regions. \texttt{cdfs-extract} is
a wrapper around the SAS tools \texttt{evselect}, \texttt{rmfgen}
and \texttt{arfgen} to extract spectra and compute the response and
ancillary matrices. The MOS and PN spectra were then summed and the
response matrices averaged using the FTOOLs \texttt{mathpha},
\texttt{addrmf} and \texttt{addarf}.

\begin{figure}
  \centering
  \includegraphics[width=\columnwidth]{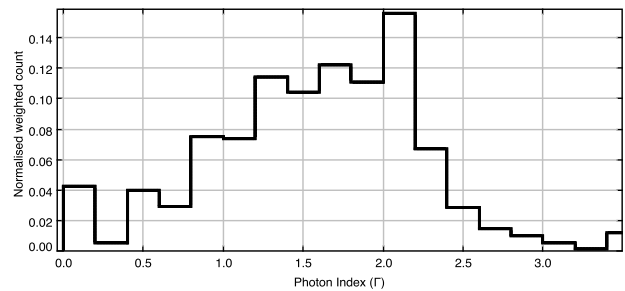}
  \caption{ Weighted normalised histogram of best-fit power-law slopes
    for the XMM-ATLAS sources. The inverse of the fit C-statistics
    were used as weights. }
  \label{fig:gammadistr}
\end{figure}

The spectra were automatically analysed with the same method and
software (automatic XSPEC fits of unbinned spectra using C-statistic)
used for the 3XMM-DR4 sources \citep{corral2014}. A simple power-law
model was fit to the data and used to derive the fluxes in the 0.5--10
and 2--10 keV bands. The fit was successful for 446
  sources\footnote{For the remaining 109 sources no constraint on the
    power-law slope could be put. These sources are on average fainter
    and detected with lower likelihood than those for which the fit
    was successful.}
The mean $\Gamma$ is $1.7\pm 0.6$, consistently with the
model assumed in Sect.~\ref{sec:catalogue}.  The histogram of best-fit
slopes is shown in Fig.~\ref{fig:gammadistr}. The presence of a number
of flat- and inverted-spectra objects can be noticed; e.g., there are
56 sources (13\% of 446) with $\Gamma\le 1.0$.

\begin{figure}
  \centering
  \includegraphics[width=\columnwidth]{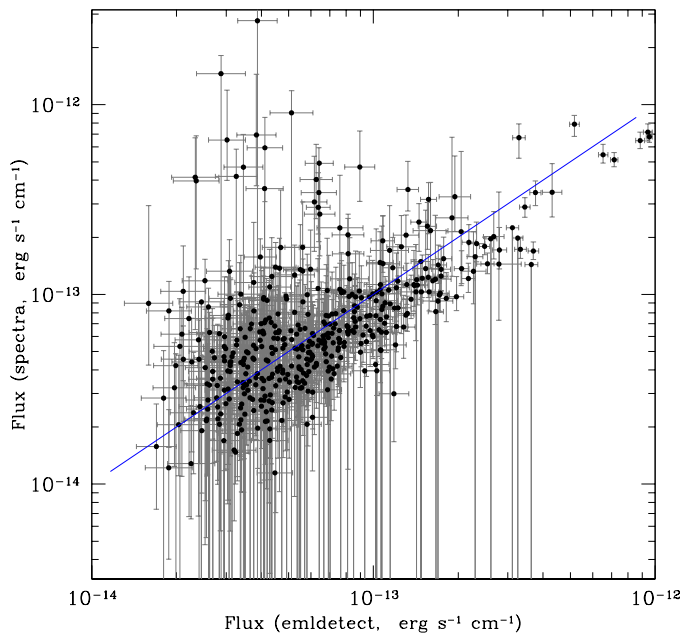}
  \caption{ Comparison of fluxes from \emld\ with fluxes from spectral
    fits. The blue line shows the 1-to-1 relationship. The fluxes are
    for the 0.5--10 keV band. }
  \label{fig:fluxflux}
\end{figure}

The comparison between the 0.5--10 keV fluxes from the spectral fits
and from the catalogue (the latter ones converted from 0.5--8 to
0.5--10 keV) is shown in Fig.~\ref{fig:fluxflux}. The distribution of
the points around the 1-to-1 line is somewhat asymmetrical for
\emld\ fluxes $\lesssim 7\e{-14}$ \ergscmq, with a larger dispersion
above the line than below; this is likely due to the combination of
the distribution of spectral slopes with the \xmm\ effective area.

A linear fit to the logarithms of fluxes shown in
Fig.~\ref{fig:fluxflux}, obtained by imposing a slope of 1, yielded a
normalisation:
\begin{equation}
F_{\rm spec} = 1\times F_{\rm cat} + 0.039 \pm 0.015
\end{equation}

corresponding to the fluxes from spectral fits being on average $(9\pm
4)\%$ brighter than those from the catalogue.  A dispersion $s=0.31$
may be given as the estimate of the standard deviation:
\begin{equation}
s = \frac{1}{N-\nu} \cdot \sqrt {\sum \left( \Log~ F_{\rm spec}-\Log~ F_{\rm cat}
\right)^2 }     \label{eqvarianza}
\end{equation}
where $\nu=1$ is the number of free parameters and $N$ is the number
of data points, $F_{\rm spec}$ is the flux from the spectral fit and
$F_{\rm cat}$ that from the catalogue.

\section{LogN-LogS}
\label{sec:lognlogs}

\begin{figure}
  \centering
  \includegraphics[width=\columnwidth]{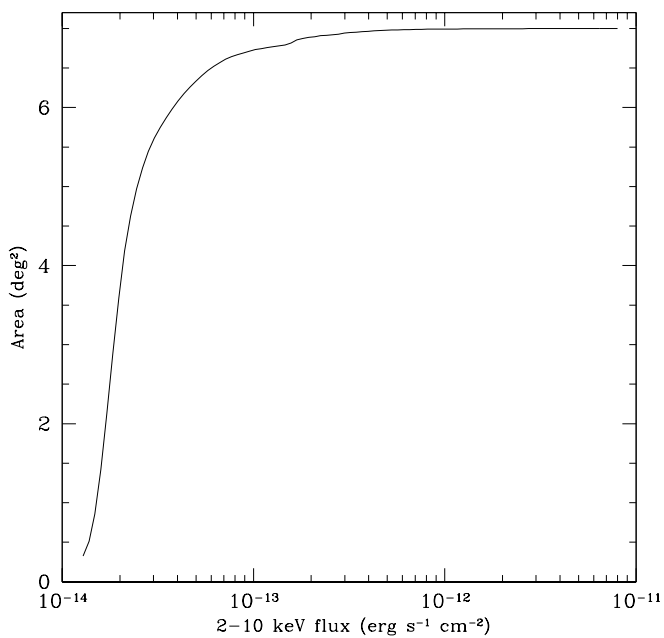}
  \caption{ Coverage in the 2-10 keV band. }
  \label{fig:coverage}
\end{figure}

The sky coverage (sky area as function of flux) for the
XMM-ATLAS survey in the 2--8 keV band was derived with the SAS
\texttt{esensmap} programme, which creates a sensitivity map by
computing count rate upper limits for each pixel. The exposure and
background maps described in Sect.~\ref{sec:detection} were used as
input. The coverage was then converted to the 2--10 keV band
(Fig.~\ref{fig:coverage}).

\begin{figure}
  \centering
  \includegraphics[width=\columnwidth]{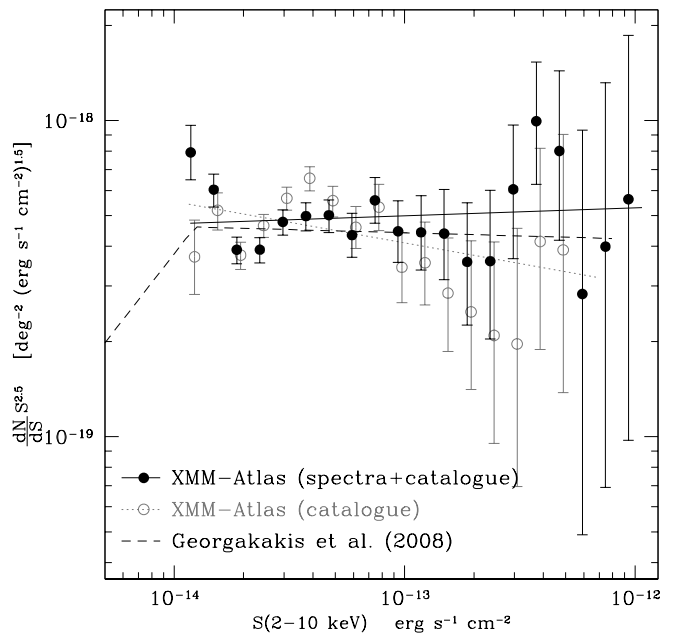}
  \caption{ XMM-ATLAS \lognlogs\ in the 2--10 keV band. Filled
    circles: fluxes from the spectral fits are used whenever possible;
    catalogue fluxes otherwise. Open circles: only fluxes from the
    catalogue are used; a small horizontal shift has been introduced
    in the plot to better distinguish the error bars.  The solid line
    shows the best-fit to the filled data points. The dashed
      line is the best-fit model from \citet{georgakakis2008}. The
      error bars show the $1\sigma$ uncertainty. }
  \label{fig:lognlogs}
\end{figure}

We compute the differential counts by binning the sources according to their
fluxes:
\begin{equation}
n(S) = \frac{1}{\Delta S} \sum_i \frac{1}{A_i}
\end{equation}
where $S$ is the central flux of the bin; $\Delta S$ is the bin width;
and for each source $i$, $A_i$ is the coverage at the source's
flux. The sum is performed on all sources with flux $S_i\in [S-\Delta
  S/2,S+\Delta S/2]$. The error on the counts in any bin can be
computed by assuming gaussianity or, for bins with less than 50 sources,
with the \citet{gehrels86} approximation.

In Fig.~\ref{fig:lognlogs} we show the 2--10 keV differential
\lognlogs\ of the ATLAS sources detected in the 2--8 keV band. Two
estimates are shown, from two different sets of fluxes:

\begin{enumerate}
\item ``spectral \lognlogs'': fluxes from the spectral fits
  (Sect.~\ref{sec:spectra}) for 387 sources with hard detection,
  spectrum, and best-fit flux larger than $10^{-14}$ \ergscmq (the
  threshold chosen for the faintest bin); plus fluxes from the catalogue
  (converted to the 2--10 keV band) for the remaining sources
  (i.e.\ those without spectra or without a successful fit);
\item ``catalogue \lognlogs'': fluxes from the catalogue for all the
  814 hard sources (also converted to the 2--10 keV band).
\end{enumerate}

It can be noticed that the spectral \lognlogs\ extends to brighter
fluxes than the catalogue one, by a factor $\lesssim 2$, which is
consistent with the use for a few sources of a $\Gamma$ flatter than
1.7.

For comparison, we also plot in Fig.~\ref{fig:lognlogs} the best-fit
model from \citet{georgakakis2008}, which was derived from a joint
analysis of several surveys, some of which are much deeper than
XMM-ATLAS. The model has a broken power-law shape, with a break flux
at $1.2\e{-14}$ \ergscmq, very close the XMM-ATLAS flux limit (at
$1.3\e{-14}$ \ergscmq\ the coverage is 5.3\% of the nominal area, and
becomes ill-defined at fainter fluxes).  Consistently, the XMM-ATLAS
\lognlogs\ does not show such a break.

A linear fit (weighted least squares) to the spectral
\lognlogs\ yielded $\Log\, N = (-2.47\pm 0.05)\Log\, S -18.0 \pm0.7$
(errors at $1\sigma$), consistent within errors with
\citet{georgakakis2008} and with a Euclidean distribution.  A fit to
the catalogue \lognlogs\ yielded $\Log\, N = (-2.37\pm 0.08)\Log\, S
-20.09 \pm1.03$, which is only consistent with Euclidean counts at a
$2\sigma$ level.

\section{Conclusions}
\label{sec:conclusion}

We have presented the observations, data reduction, catalogue and
number counts of the XMM-ATLAS survey, which covers 7.1 deg$^2$ with
flux limits (defined as the flux of the faintest detected sources) of
$2\e{-15}$, $6\e{-15}$ and $9\e{-15}$ \ergscmq\ in the 0.5--2, 0.5--8
and 2--8 keV bands, respectively.

We derived the catalogues with a two-step procedure. First, the
\ewav\ SAS task was used to identify candidate sources with a
significance equivalent to $4\sigma$, and to find their
coordinates. Next, we used the SAS \emld\ tool to further check the
significance of the sources, and obtain counts, count rates and
fluxes. The final catalogues contain 1700, 1582 and 814 sources in the
0.5--2, 0.5--8 and 2--8 keV bands, respectively, with a total of 1816
unique sources.  A list of supplementary sources, detected by
\ewav\ but not confirmed by \emld, is also provided.

To investigate the effect of assuming a common spectral model for all
sources to convert count rates to fluxes, we extracted spectra for
each source with at least 40 counts and fitted them with a power-law
model. We found that on average, the fluxes from the spectral
  fits are $(9 \pm 4)\%$ brighter than those from assuming a common
power-law photon index of $\Gamma=1.7$. Also, the average best-fit
spectral slope is $\Gamma=1.7\pm 0.6$.

We derived the 2--10 keV differential \lognlogs\ for the XMM-ATLAS
sources, which spans the $10^{-14}$--$10^{-12}$ \ergscmq\ flux
interval. Using the fluxes from the spectral fits (for sources with
spectra; and fluxes from the catalogue for all other sources) produces
a \lognlogs\ which is less noisy and more consistent with Euclidean
counts than using fluxes from the catalogue. A weighted linear fit
yielded $\Log\, N = (-2.47\pm 0.05)\Log\, S -18.0 \pm0.7$
(errors at $1\sigma$); this is consistent within errors with previous
studies, e.g.\ \citet{georgakakis2008}.

Finally, we release the software tools which have been developed or
enhanced to accomplish the analysis in this paper: {\tt griddetect}, a
programme to run \emld\ over a very wide mosaic; {\tt autoregions}, to
define extraction regions for spectra and aperture photometry; and
{\tt cdfs-extract}, to extract spectra for multiple sources from
multiple observations.

\begin{appendix} 
\section{Released software}
\label{sec:software}

For the software described below, more information and instructions
for use can be found on the author's
website\footnote{\texttt{members.noa.gr/piero.ranalli}} and source
repository\footnote{\texttt{github.com/piero-ranalli}}.

\subsection*{{\tt cdfs-extract} and {\tt autoregions}}

Given a number of event files and a list of source and background
positions, the {\tt cdfs-extract} programme checks if the source and
background are in the field of view of any observation, and it
extracts products accordingly. The observations may or may not
overlap. 

The extracted products are spectra (source and background) and
responses (source and, optionally, background RMFs and ARFs), or
aperture photometry, or lightcurves.

The programme takes as input two text files, containing:
\begin{itemize}
\item a list of event files, exposure maps and (optional) images;
\item a list of source and background positions and radii (the camera
  may optionally be specified, allowing the same source to have
  different positions/radii for different cameras).
\end{itemize}

Spectra and responses can be summed if the user wants, producing a
single spectral file for each source and each camera.

The list of source and background positions can be either be manually
written, or automatically produced by {\tt autoregions}.  If the
sources are not too close to each other and the background has no
large variations on scales $\lesssim 20$--$30\arcsec$ then annuli can
be used as background regions and their positions and radii can be
automatically generated (i.e., the user is dealing with a shallow
survey such as XMM-ATLAS, XXL or XMM-COSMOS; conversely, deep surveys
such as the XMM-CDFS feature close groups of sources and complex
spatial patterns in the background which make automatic procedures
unreliable). 

Given a catalogue of sources, stored in a database (e.g.: PostgreSQL)
and containing source counts and background surface brightnesses
(i.e., the \emld\ output), the {\tt autoregions} programme computes
extraction regions by maximizing the expected signal/noise
ratio. Overlapping regions are identified, and where possible, the
overlapping areas are excised.

Besides this work, the programmes described above have been, or are
being used, in several papers, e.g.\ the obscured AGN in XMM-COSMOS
\citep{lanzuisi2015}, the XMM-CDFS spectral survey (Comastri et al.,
in prep.), the XXL brightest AGN spectral survey (Fotopoulou et al.,
in prep.).

Both {\tt cdfs-extract} and {\tt autoregions} are written in
(modern-style) Perl using the Perl Data Language (PDL) libraries, and
are free software released under the terms of the GNU Affero GPL
license\footnote{The Affero GPL is a variant of the
  GPL which also adds protection when the software is used as a web
  service. \label{fn:AGPL}}.

\subsection*{{\tt griddetect} and associated libraries}

The motivation for {\tt griddetect}, and how it works, have already
been described in Sect.~\ref{sec:whygriddetect} (see also
Fig.~\ref{fig:detectiongrid}). 

Here we also mention that the \xmm\ SAS provides a tool
(\texttt{emosaicproc}), whose functionality is partly duplicated by
{\tt griddetect}. The main difference between the tools may be
summarized as follows: \texttt{emosaicproc} automatically computes the
grid, and requires less data preparation by the user if a single
mosaic-mode obsid is to be processed, but only works on a
  single obsid; {\tt griddetect} needs the user to define the grid
cracks, but it is easier to add pointings from different obsids, and
it can also compute background maps.

{\tt griddetect} works on individual pointings, splitted from
mosaic-mode event files;
it also computes background maps according to the
\citet{xmm-cosmos} method. The input comprises:
\begin{itemize}
\item the grid specification: a rotation angle, and the x and y
  coordinates (rotated RA and DEC) of the grid lines;
\item the list of pointings with their coordinates.
\end{itemize}
The output is a series of two catalogues per detection cell: one
containing the detected sources, and another one containing only the
sources within $3\arcsec$ from the cracks (used to check if there are
missing or repeated sources, arising because of numeric effects
placing them alternatively on the two sides of the crack).

Together with {\tt griddetect}, we also release a set of associated
libraries which may be of more general purpose: {\tt XMMSAS::Extract}
and {\tt XMMSAS::Detect}. These are object-oriented Perl packages
which present a high-level interface to some SAS and FTOOLS
commands. They can be called by Perl programmes to filter event files and extract
images and exposure maps using either pre-defined filters (e.g., the
instrumental lines excluded in Sect.~\ref{sec:observations}) or
user-defined ones; and to compute background maps, source masks and
call \emld.

{\tt griddetect} and its associated libraries are written in
(modern-style) Perl using the Moose object system and the Perl Data
Language (PDL) libraries, and are free software released under the
terms of the GNU Affero GPL license.

\end{appendix}

\begin{acknowledgements}
We thank an anonymous referee whose comments have contributed to
improve the presentation of this paper. We thank N. Cappelluti for
valuable discussions about \emld.  PR acknowledges a grant from the
Greek General Secretariat of Research and Technology in the framework
of the programme Support of Postdoctoral Researchers. ADM acknowledges
financial support from the UK Science and Technology Facilities
Council (ST/I001573/I). The use of Virtual Observatory tools is
acknowledged (TOPCAT, \citealt{topcat}, and the Aladin sky atlas,
\citealt{aladin}).

\end{acknowledgements}

\bibliographystyle{aa}
\bibliography{../fullbiblio}

\end{document}